# All-Temperature Superconductor

Ron Bourgoin
Edgecombe Community College
Rocky Mount, North Carolina 27801
bourgoinr@edgecombe.edu

**Abstract**
Several surprises are beginning to emerge from studies of nanostructures: whereas increased resistance with decreased thickness has been expected, the exact opposite in several instances has been found. It is beginning to seem we are in the process of wedging open the way to a superconductor that will operate at least at room temperature. This paper will examine some pertinent experimental findings since the turn of the century.

In the September, 2005 issue of Nikkei Electronics magazine, my article appeared about the extremely high conductivity at room temperature of nanofilaments I had made from chemical solutions. By inserting oppositely-directed pin electrodes in a solution of metal nanoparticles suspended in hot epoxy resin, a voltage applied across the separated pin tips caused nanofilaments to be formed. When the epoxy hardened, permanent filaments of length 1 $mm$ were formed with ends fused to the pin tips. Helical filaments up to 4 $mm$ were obtained by retracting one of the pins.[1] In addition to high conductivity, the filaments had the surprising ability to continue conducting through the ohmic network even when the power was shut off. This persistent current endured for up to 5 seconds. Another surprising finding was that when the filaments were cycled, resistance fell during each cycle until, after six cycles, resistance was zero. It seems that the electrons in the persistent current going one way are pairing with electrons in the AC current going the other way. A typical sample with 15 initial ohms drops nearly 3 ohms/cycle. Understanding what's going on has been quite a challenge, and the following discussion should be construed as an attempt to make sense of the experimental findings.

A superconducting unit of a pair of electrons that keeps moving through an ohmic circuit when the power is shut off has to be quite small. We begin the analysis with the concept of two electrons rotating about a common center of mass so we can get an idea of the size of the pair. From

$$m_e r u = \tfrac{1}{2}\hbar \tag{1}$$

we obtain the attractive force acting between the electrons,

$$F_a = \frac{(\tfrac{1}{2}\hbar)^2}{m_e r^3} \tag{2}$$

Equating this force with that of electrical repulsion for balance yields a distance of separation of $r = 1.325 \times 10^{-11} m.$ This is one-eighth the size of the ground-state hydrogen atom.

There is another force at work, which is the magnetic dipolar force, given by

$$F_m = \frac{k\mu_B^2}{r^4} \tag{3}$$

which prevents the pair from achieving too close proximity and crushing itself out of existence. The value of $k$ is $4.66 \times 10^{-4} kg \cdot m/coul^2$, and $\mu_B$ is the Bohr magneton.

I am encouraged to see Beijing National Laboratory report that ultrathin films of 26 atoms thickness have half the ohmic resistance of thicker films 28 atoms thick; films 21 atoms thick are 20% less resistive than the larger films 25 atoms thick. One of the most dramatic falls in resistance occurs from the 27-atom-thick to the 26-atom-thick film: there is a 60% drop in resistance by shaving off a single monolayer. [2,3]

Equally gratifying was the recent article from the National Laboratory at Nanjing, which reminds us all that sets of electrons in atomic orbitals have been superconducting since the beginning of time, and if we artificially create the size of an atomic orbital, we should be able to attain superconductivity at any temperature. [4] It is believed the ultrathin wire will be , in essence, a virtual atomic orbit. If we use Beijing lab's figure for the electron wavelength, $\lambda_e = 1.06\ nm$, and Nanjing lab's figure for the radius of the atomic orbital, we obtain

$$r = \frac{\lambda_e}{2\pi} \tag{4}$$

for the radius of the wire that should provide us with superconductivity at 300K. It is rather interesting to observe that this resembles the equation for the radius of a waveguide, which is what the nanowire, in essence, would be. Equation (4) predicts a nanofilament diameter of one-third nanometer, or two atoms. If the Beijing lab, with capability to lay down single-atom layers, will lay down two layers two atoms wide, for a total of four atoms in the cross-section, it will see a room-temperature superconductor. This conductor should superconduct on both DC and AC systems.

Although quantum size effects have been expected to result in increased resistance in nanostructures of widths under 10 $nm$, we are discovering several surprises. When relatively large gold leads to molecular wires are replaced by single atoms, for instance, resistance goes down, contrary to expectations.[5] In graphene sheets of a single atom thickness, for another example, high conductivity with electron drift velocities of $10^6$ $m/\sec$ have been measured.[6,7] The same can be said of 2D quantum dots under certain conditions of temperature.[8] Those pesky dendritic carbon arborizations in electrical shorts that have enormous current ampacities might now be perceived in a new light: perhaps they have been room-temperature superconductors right under our noses.

Pertinent to this discussion is the fact that the human central nervous system is laced everywhere with carbon nanofibers. No-one has understood their real function. Perhaps the reason we have not been able to repair the injured spinal cord is because we have not ensured continuity of the carbon filaments. This statement is not meant in any way to diminish the gargantuan efforts over the past 50 years to regenerate CNS spinal nerves; it is, however, a call to examine a possibility that is emerging from discoveries in nanoscience. What sort of conductivity can we expect from a carbon nanofilament? To keep it simple, we shall assume a single row of carbon atoms of one centimeter length. The applicable equation is

$$I = neu \qquad (5)$$

which a rough, order-of-magnitude calculation yields for the current, $I = 10 \ \mu A$, which is definitely in the range of axonal currents.[9]

For those who are interested in highly conductive nanostructures to produce magnetic fields at 300K, a bundle consisting of one hundred ambient-temperature superconductive nanothreads is expected to produce a magnetic field of

$$B = \frac{\mu_0 I}{r} = 12.56T \qquad (6)$$

Considering that most of the present cost of a magnetic resonance image (MRI) is for the cryogenic generation of a magnetic field, the offspring from the marriage of nanotechnology and superconductivity, which will eliminate the cryogenics, will be welcomed.

In conclusion, we have examined several instances of extremely high conductivity and low resistance at 300K in nanowires, graphene sheets, ultrathin films, single-atom leads to molecular wires, and quantum dots. Some suggestions were offered that the carbon nanofibers found in the human spinal cord participate in bioconduction, and that cable made from highly conductive nanowire can provide intense magnetic fields. A non-BCS mechanism of Cooper-pair formation was offered based on recent theoretical work in China. If research activities continue at their present pace, it is expected a room-temperature superconductor will be found before the end of this decade.

I wish to thank Xiuqing Huang for excellent advice and information.